\documentclass[a4paper,11pt,twoside]{article}
\usepackage[INRIA]{lipinria}
\usepackage[latin1]{inputenc}
\usepackage{a4wide}
\usepackage{amsfonts,amsmath,amssymb,amsthm,amstext,latexsym,paralist}
\usepackage{url}
\usepackage{xspace}
\usepackage{algorithm2e}
\usepackage{subfigure}
\usepackage{figlatex}
\makeatletter
\newcommand{\eqnlabel}[1]{\xdef\@currentlabel{\theequation}\ltx@label{#1}}
\newcommand{\cnstslabel}[2][\thesubcounstraints]{
  \refstepcounter{subcounstraints}%
  \xdef\@currentlabel{#1}%
  \tagform@{#1}%
  \ltx@label{#2}\quad &}
\newcommand{\cnstlabelBIS}[2][\thesubcounstraints]{
  \xdef\@currentlabel{#1}%
  \ltx@label{#2}}
\newcounter{subcounstraints}[equation]

\newenvironment{LinearProgram}[2][Minimize]{%
  \begin{equation}%
    \let\label=\eqnlabel
    \begin{array}{l}%
      \textsc{#1 } #2, \\%
      \textsc{under the constraint}\\%
    \end{array}
  \end{equation}%
}

\newcommand{\EquationsNumbered}[1]{%
  \setcounter{subcounstraints}{0}
  \renewcommand{\thesubcounstraints}{\theequation\alph{subcounstraints}}
  \def\set@counstraintscounter{
    \refstepcounter{subcounstraints}%
    \xdef\@currentlabel{\thesubcounstraints}%
    \tagform@{\thesubcounstraints}%
  }
  \def\mark{\set@counstraintscounter\quad&}
  \def\n{\\\mark}
  \begin{aligned}%
    #1
  \end{aligned}%
}
\makeatother

\newtheorem{lemma}{Lemma}
\newtheorem{theorem}{Theorem}

\AtBeginDocument{}

\newcommand{\comm}{\textsf{$\delta$}} % comm for stage
\newcommand{\calc}{\textsf{w}}    % calc for  stage
\newcommand{\speed}{\textsf{s}}   % speed of proc
\newcommand{\bw}{\textsf{b}}      % bandwidth of link

\newcommand{\alloc}{\textsf{alloc}}

\newcommand{\latency}{T_{\textsf{latency}}}

\newcommand{\FULLHOM}{\textit{Fully Homogeneous}\xspace}
\newcommand{\COMHOM}{\textit{Communication Homogeneous}\xspace}
\newcommand{\HETERO}{\textit{Fully Heterogeneous}\xspace}
\newcommand{\FPHOM}{\textit{Failure Homogeneous}\xspace}
\newcommand{\FPHET}{\textit{Failure Heterogeneous}\xspace}

\newcommand{\n}{n} %\textsf{n}} % number of stages
\newcommand{\m}{m} %\textsf{p}} % number of procs
\newcommand{\fp}{\textsf{fp}} %\textsf{p}} % number of procs
\newcommand{\inn}{\textsf{in}}
\newcommand{\out}{\textsf{out}}

\newcommand{\link}{\textsf{link}} % number of procs
\renewcommand{\S}{\mathcal{S}} % stage
\renewcommand{\L}{\mathcal{L}} % latency
\newcommand{\FP}{\mathcal{FP}} % global failure probability
\newcommand{\II}{\mathcal{I}} % for the proofs
\newcommand{\Inst}{\mathcal{I}} % for the proofs

\begin{document}

\RRIthead{Optimizing Latency and Reliability
of Pipeline Workflow Applications}

\RRIkeywords{Heterogeneity, scheduling, complexity results, reliability, response time.}
\RRImotscles{H\'et\'erog\'en\'eit\'e, ordonnancement, r\'esultats de
  complexit\'e, fiabilit\'e, temps de r\'eponse.}

\RRIdate{
March 2008}

\RRIabstract{Mapping applications onto heterogeneous platforms is a
difficult challenge, even for simple application patterns such as
pipeline graphs. The problem is even more complex when processors are
subject to failure during the execution of the application.

In this paper, we study the complexity of a bi-criteria mapping which
aims at optimizing the latency ({\em i.e.}, the response time)  and the
reliability ({\em i.e.}, the probability that the computation will be
successful) of the application. Latency is minimized by using faster
processors, while reliability is increased by replicating
computations on a set of processors. However, replication increases
latency (additional communications, slower processors). The application
fails to be executed only if all the processors fail during execution.

While simple polynomial
algorithms can be found for fully homogeneous platforms, the problem
becomes NP-hard when tackling heterogeneous platforms. This is yet another illustration
of the additional complexity added by heterogeneity.
}

\RRIresume{L'ordonnancement et l'allocation des applications sur plates-formes
h\'et\'erog\`enes sont des probl\`emes cruciaux, m\^eme pour des
applications simples comme des graphes en pipeline. Le probl\`eme
devient m\^eme encore plus complexe quand les processeurs peuvent
tomber en panne pendant l'ex\'ecution de l'application. Dans cet article,
nous \'etudions la complexit\'e d'une allocation bi-crit\`ere qui vise
\`a optimiser la latence (i.e., le temps de r\'eponse) et la fiabilit\'e
(i.e., la probabilit\'e que le calcul r\'eussisse) de l'application. La
latence est minimis\'ee en utilisant des processeurs rapides, tandis
que la fiabilit\'e est augment\'ee en r\'epliquant les calculs sur un
ensemble de processeurs. Toutefois, la r\'eplication augmente la latence
(communications additionnelles et processeurs moins
rapides). L'application \'echoue \`a \^etre ex\'ecut\'ee seulement si
tout les processeurs \'echouent pendant l'ex\'ecution. Des
algorithmes simples en temps polynomial peuvent \^etre trouv\'es pour
plates-formes compl\`etement homog\`enes, tandis que le probl\`eme
devient NP-dur quand on s'attaque aux plates-formes h\'et\'erog\`enes. C'est
encore une autre illustration de la complexit\'e additionelle due \`a
l'h\'et\'erog\'en\'eit\'e.}

\RRIahead{A. Benoit \and V. Rehn-Sonigo \and Y. Robert}
\RRIauthor{Anne Benoit \and Veronika Rehn-Sonigo \and Yves Robert}

\RRItitle{Optimizing Latency and Reliability
of Pipeline Workflow Applications}
\RRItitre{Optimisation de latence et fiabilit\'e des applications de
  type workflow pipelin\'e}

\RRInumber{RR2007-43}
\RRItheme{\THNum}
\RRIprojet{GRAAL}
\RRNo{6345}
\RRInumber{6345} 
\RRImaketitle

\section{Introduction}

Mapping applications onto parallel platforms is a difficult
challenge. Several scheduling
and load-balancing techniques have been developed for homogeneous architectures
(see~\cite{ieee-sched} for a survey) but the advent of heterogeneous
clusters has rendered the mapping problem even more
difficult.
Moreover, in a distributed computing architecture, some processors may
suddenly become unavailable, and we are facing the problem of
failure~\cite{abawajy04,albers01}.
In this context of dynamic heterogeneous platforms with failures, a
structured programming
approach rules out many of the problems which the
low-level parallel application developer is usually confronted to,
such as deadlocks or process starvation.
%Moreover, many real applications draw from a range of well-known
%solution paradigms, such as pipelined or farmed
%computations. High-level approaches based on algorithmic
%skeletons~\cite{Co02,Ra02} identify such patterns and seek to make it
%easy for an application developer to tailor such a paradigm to a
%specific problem. A library of skeletons is provided to the programmer,
%who can rely on these already coded patterns to express the
%communication scheme within its own application.
%Moreover, the use of a particular skeleton carries with it considerable
%information about implied scheduling dependencies, which we believe
%can help address the complex problem of mapping a distributed
%application onto a heterogeneous platform.

In this paper, we consider %regular
application workflows that can be expressed
%as algorithmic skeletons, and we focus on the pipeline skeleton, which
as pipeline graphs. Typical applications include digital image
processing, where images have to be processed in steady-state mode. A
well known pipeline application of this type is for example JPEG
encoding (see http://www.jpeg.org/).
 %, since it is one of the most widely used.
In such workflow applications,
a series of data sets (tasks) enter the
input stage and progress from stage to stage until the final result is
computed.
Each stage has its own communication and computation requirements: it
reads an input file from the previous stage, processes the data and
outputs a result to the next stage.
For each data set, initial data is input to the first stage, and final
results are output
from the last stage. %The pipeline workflow operates in synchronous
%mode: after some latency due to the initialization delay, a new task
%is completed every period. %The period is defined
%as the longest cycle-time to operate a stage. %, and is the inverse of
%the throughput that can be achieved.

Each processor has a failure probability, which expresses the chance
that the processor fails during execution. Key metrics for a
given workflow are the latency and the failure probability.
The latency is the time elapsed between the beginning and the end of
the execution of a given data set, hence it measures the response time
of the system to process the data set entirely.
Intuitively, we minimize the latency by assigning all stages to the
fastest processor, but this may lead to an unreliable execution of the
application. Therefore, we need to find trade-offs between two
antagonistic objectives, namely latency and failure probability.
Informally, the application will be reliable for a given mapping if
the corresponding global failure probability is small.
Here, we focus on bi-criteria approaches, i.e.,
minimizing the latency under failure probability constraints, or the
converse. Indeed, such bi-criteria approaches seem more natural than
the minimization of a linear combination of both criteria. Users
may have latency constraints or reliability constraints, but it makes
little sense for them
to minimize the sum of the latency and of the failure probability.

%The problem of mapping pipeline skeletons onto parallel platforms has
%received some attention,
%and we survey related work in Section~\ref{sec.conc}.
%In this
%paper, we target heterogeneous platforms, and aim at deriving optimal
%mappings for a bi-criteria objective function.
%Each pipeline stage can be seen as a sequential procedure which may
%perform disc accesses or write data in the memory for each task. This
%data may be reused from one task to another, and thus the rule of
%the game is always to process the tasks in a sequential order within a
%stage. Moreover,
%Due to the possible local memory accesses, a given
%stage must be mapped onto a single processor: we cannot process half
%of the tasks on a processor and the remaining tasks on another without
%exchanging intra-stage information, which might be costly and
%difficult to implement.
%
We focus on pipeline skeletons and thus
we enforce the rule that a given stage is mapped onto a single processor.
In other words, a processor
that is assigned a stage will execute the operations required by this
stage (input, computation and output) for all the tasks fed into the
pipeline.
However, in order to improve reliability, we can replicate the
computations for a given stage on several processors, i.e., a set of
processors performs identical computations on every data set. Thus, in
case of failure, we can take the result from a processor which is
still working.
The optimization problem can be stated informally as
follows: which stage to assign to which (set of) processors? We
require the mapping to be interval-based, i.e., a set of processors is
assigned an interval of consecutive stages.
The main objective of this paper is to assess the complexity of this
bi-criteria mapping problem.

The rest of the paper is organized as follows.
Section~\ref{sec.model} is devoted to the presentation of the
target optimization problems. Next in Section~\ref{sec:examples}
some motivating examples are presented. In Section \ref{sec:complexity} we
proceed to the complexity results.
Finally, we briefly review related work and state some concluding remarks
in Section~\ref{sec.conc}.

\section{Framework and optimization problems}
\label{sec.model}

\subsection{Framework}

\begin{figure}[tbh]
\centering
\includegraphics[width=0.5\textwidth]{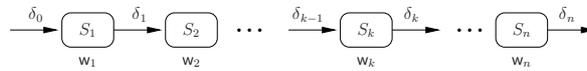}
%\vspace{-0.5cm}
\caption{The application pipeline.}
\label{fig.pipeline}
\end{figure}

The application is expressed as a pipeline graph of $\n$ stages
$\S_k$, $1 \leq k \leq \n$, as illustrated on
Figure~\ref{fig.pipeline}. Consecutive data sets are fed into the
pipeline and processed from stage to stage, until they exit the
pipeline after the last stage.
Each stage executes a task. More precisely, the $k$-th stage $\S_k$
receives an input from the previous stage, of size $\comm_{k-1}$, performs a number of $\calc_k$ computations, and outputs data of size $\comm_{k}$ to the
next stage. This operation corresponds to the $k$-th task and is
repeated periodically on each data set.
The first stage $\S_1$ receives an input of size $\comm_0$ from the outside
world, while the last stage $\S_{\n}$ returns the result, of size
$\comm_{\n}$, to the outside world.

\begin{figure}[tbh]
\begin{center}
\includegraphics[height=4cm]{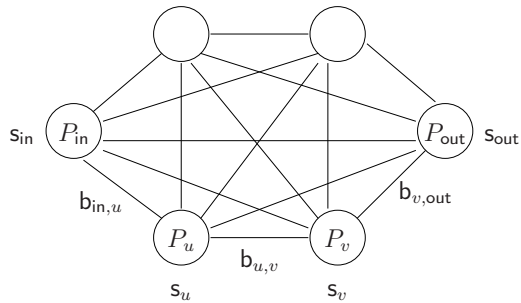}
\end{center}
\caption{The target platform.} \label{fig.clique}
\end{figure}

We target a platform (see Figure~\ref{fig.clique}),
with $\m$ processors $P_u$, $1 \leq u \leq \m$, fully interconnected
as a (virtual) clique.
We associate to each processor a failure probability $0\leq \fp_u \leq
1$, $1 \leq u
\leq \m$, which is the probability that the processor breaks down
during the execution of the application. %, to crash down.
A set of processors with identical failure probabilities is denoted
\FPHOM and otherwise \FPHET.
We consider a constant failure probability as we are dealing with
workflows. These workflows are meant to run during a very long time,
%, a priori, do not have a beginning or end
and therefore we address the question of whether the processor
will break down or not at any time during execution. Indeed the maximum
latency will be determined by the latency of the datasets which are
processed after the failure.

There is a bidirectional link $\link_{u,v}: P_u \to P_v$ between any
processor pair $P_u$ and $P_v$,
of bandwidth $\bw_{u,v}$. %Note that we do not need to have a physical link between any processor pair. Instead, we may have a switch, or even a path composed of several physical links, to interconnect $P_u$ and $P_v$; in the latter case we would retain the bandwidth of the slowest link in the path for the value of $\bw_{u,v}$.
The speed of processor $P_u$ is denoted as
$\speed_u$, and it takes $X/\speed_u$ time-units for $P_u$ to execute $X$ floating point
operations. We also enforce a linear cost model for communications,
hence it takes
$X/\bw_{u,v}$ time-units to send (or receive) a message of size $X$ from $P_u$ to $P_v$.
Communication contention is taken care of by enforcing the \emph{one-port}
model~\cite{Bhat99efficient,BhatRagra03}. In this model, a given
processor can be involved in a single communication at any time-step, either a send or a receive. However,
independent communications between distinct processor pairs can take place
simultaneously. The one-port model seems to fit the performance of
some current MPI implementations, which serialize asynchronous MPI
sends as soon as message sizes exceed a few megabytes~\cite{SaifPa2004}.
%We detail the communication model between replicated sets of
%processors while discussing the mapping problem.

We consider three types of platforms:
\begin{itemize}
\item \FULLHOM platforms have
identical processors ($\speed_u=\speed$ for $1\leq u \leq \m$) and
interconnection links ($\bw_{u,v}=\bw$ for $1\leq u,v \leq \m$);
\item \COMHOM platforms, with identical links but different speed
processors, introduce a first degree of
heterogeneity;
\item \HETERO platforms constitute the most difficult instance,
with different speed processors and different capacity links.
\end{itemize}

Finally, we assume that two special additional processors $P_{\inn}$
and $P_{\out}$ are devoted to input/output data. Initially, the input
data for each task resides on $P_{\inn}$, while all results must be
returned to and stored in $P_{\out}$.

%\medskip

\subsection{Bi-criteria Mapping Problem}
The general mapping problem
consists in assigning application stages
to platform processors. For simplicity, we could assume that
each stage $\S_i$ of the application
pipeline is mapped onto a distinct processor (which is possible only if
$\n \leq \m$).
However, such one-to-one mappings may be unduly restrictive, and a
natural extension is to search for interval mappings, i.e., allocation
functions where each participating processor is assigned an interval
of consecutive stages. Intuitively, assigning several consecutive
tasks to the same processor will increase its computational load,
but may well dramatically decrease communication requirements.
In fact, the best
interval mapping may turn out to be a one-to-one mapping, or instead
may enroll only a very small number of fast computing processors
interconnected by high-speed links.
Interval mappings constitute a natural and useful generalization of
one-to-one mappings (not to speak of situations where $\m < \n$, where
interval mappings are mandatory), and such mappings have been studied
by Subhlock et al.~\cite{subhlock-ppopp95,subhlock-spaa96}.

%parler de one-to-one et interval=generalization
%
%It is natural to map
%intervals of consecutive stages onto processors~\cite{subhlock-ppopp95,subhlock-spaa96}.
%Intuitively, assigning several consecutive
%tasks to the same processor will increase their computational load, but may well
%dramatically decrease communication requirements.
%The cost model associated to
%interval mappings is the following. W
Formally, we search for a partition of
$[1..\n]$ into $p \leq \m$
intervals $I_j = [d_j, e_j]$ such that $d_j \leq e_j$ for $1 \leq j
\leq p$, $d_1 = 1$,
$d_{j+1}= e_j + 1$ for $1 \leq j \leq p-1$ and $e_p = \n$.

%\subsection{Replication: A mechanism to increase fiability}
%Processors have a failure probability $fp_u$, $1 \leq u \leq \m$, to
% crash down. To increase the probability that the entire pipeline of
% stages is processed, it is  possible to map stages or intervals of
% stages on multiple processors.
The function $\alloc(j)$ returns the indices of the processors on
which interval $I_j$ is mapped. There are $k_j = |\alloc(j)|$
processors executing $I_j$,
and obviously $k_j \geq 1$. Increasing $k_j$ increases the reliability
of the execution of interval $I_j$.
The optimization problem is to determine the best mapping, over all possible
partitions into intervals, and over all processor assignments. The
objective can be to minimize either the latency or the failure
probability,  or a combination: %given a threshold
%period, what is the minimum latency that can be achieved? and the counterpart:
%given a threshold latency, what is the minimum period that can be
%achieved? in this paper we focus on the combination latency - failure
%probability:
given a threshold latency,  what is the minimum failure probability
that can be achieved? Similarly, given a threshold failure probability, what is the
minimum latency  that can be achieved?

%\paragraph{Fiability}

The failure probability can be computed given the number $p$ of intervals
and  the set of processors assigned to each interval:
%\begin{equation}
$\FP = 1 - \prod_{1\leq j \leq p}(1 - \prod_{u\in \alloc(j)}{\fp_{u}})$.
%\end{equation}

%\subsection{Influence of platform parameters}
%We consider different types of platforms,
%Depending on platform parameters, the treated target platforms will
%have different characteristics: The most restricted platform type is
%called \FULLHOM, where all processors have the same speed and the
%interconnection links have the same capacities. \COMHOM platforms
%consist in different speed processors that are ($s_u \neq s_v$) interconnected
%by links of the same bandwidth. In the most general case, we have
%\HETERO platforms, where different speed processors are interconnected
%via links of different capacities.

%In the following we present the formal descritptions of latency and
%failure probability under the differnet platform models.

%\subsection{The different platform constellations}
%\paragraph{Latency on \COMHOM and \FULLHOM  platforms}

%, $1 \leq z \leq k_j$, and the period is expressed as
%% Dealing with \FULLHOM or \COMHOM platforms, the period is expressed as

%% \begin{equation}
%% \label{eq.interval.period}
%% \period = \max_{1 \leq j \leq p} \left\{  \frac{k_{j-1}\times \comm_{d_j -
%% 1}}{\bw}%_{\alloc(j-1),\alloc(j)}}
%% + \frac{\sum_{i=d_j}^{e_j}
%% \calc_{i}}{\min_z(\speed_{\alloc(j,z)})} +
%% \frac{k_{j+1}\times \comm_{e_j}}{\bw}%_{\alloc(j),\alloc(j+1)}}
%% \right\},
%% \end{equation}
%% where $k_0 = k_{p+1}=1$.

%\noindent
We assume that $\alloc(0) = \{\inn\}$ and $\alloc(m+1) = \{\out\}$,
where $P_{\inn}$ is a special processor holding the initial data, and
$P_{\out}$ is receiving the results.
Dealing with \FULLHOM and \COMHOM platforms, the latency
is obtained as
%by the following expression:
% (data sets traverse all
%stages, and only interprocessor communications need to be paid for):

\begin{equation}
\label{eq.interval.latency}
%\small{
\latency = \sum_{1 \leq j \leq p} \left\{ k_j\times \frac{\comm_{d_j -
1}}{\bw}%_{\alloc(j-1),\alloc(j)}}
+ \frac{\sum_{i=d_j}^{e_j}
\calc_{i}}{\min_{u\in \alloc(j)}(\speed_u)}
\right\} + \frac{\comm_n}{\bw}.%}
\end{equation}

In equation~(1), we consider the longest path required to compute a given data
set. The worst case is when the first
processors involved in the replication  fail during execution.
A communication to interval~$j$ must then be paid $k_j$ times since these
are serialized (one-port model). For computations, we consider the
total computation time required by the slowest processor assigned to the interval.
For the final output, only one communication is required, hence
the~$\comm_n/\bw$.
Note that in order to achieve this latency, we need a
standard consensus protocol to determine which of the
surviving processors performs the outgoing communications
\cite{Tel2000}.

A similar mechanism is used
for \HETERO platforms:
\begin{equation}
\label{eq.interval.latency.het}
%\small{\begin{split}
\latency = \sum_{u\in\alloc(1)} \frac{\comm_0}{\bw_{\inn,u}} +
\sum_{1 \leq j \leq p}   \max_{u\in \alloc(j)} \left\{
\frac{\sum_{i=d_j}^{e_j} {\calc_i}}{\speed_u} + \sum_{v\in\alloc(j+1)}
\frac{\comm_{e_j}}{\bw_{u,v}}
\right\}
%\end{split}}
\end{equation}

%\paragraph{Latency on \HETERO platforms}

%, $1 \leq z \leq k_j$, and the period is expressed as
%% Dealing with \FULLHOM or \COMHOM platforms, the period is expressed as

%% \begin{equation}
%% \label{eq.interval.period}
%% \period = \max_{1 \leq j \leq p} \left\{  \frac{k_{j-1}\times \comm_{d_j -
%% 1}}{\bw}%_{\alloc(j-1),\alloc(j)}}
%% + \frac{\sum_{i=d_j}^{e_j}
%% \calc_{i}}{\min_z(\speed_{\alloc(j,z)})} +
%% \frac{k_{j+1}\times \comm_{e_j}}{\bw}%_{\alloc(j),\alloc(j+1)}}
%% \right\},
%% \end{equation}
%% where $k_0 = k_{p+1}=1$.

%\noindent %Here, we assume that $\alloc(0) = \inn$ and $\alloc(m+1) = \out$.
%On \HETERO platforms, the axpression of the latency changes (without the
%additional costs for synchronization) :

%% \begin{equation}
%% \label{eq.interval.latency}
%% \latency = \sum_{1 \leq j \leq k_0} \comm_0 + \sum_{1\leq j \leq p} \max_j\left(\sum\sum_{i=d_{j}}^{e_j}\calc_j \left\{ k_j\times \frac{\comm_{d_j -
%% 1}}{\bw}%_{\alloc(j-1),\alloc(j)}}
%% + \frac{\sum_{i=d_j}^{e_j}
%% \calc_{i}}{\min_{z}(\speed_{\alloc(j,z)})}
%% \right\} + \comm_p
%% \end{equation}

\section{Motivating examples}
\label{sec:examples}

Before presenting complexity results in Section~\ref{sec:complexity},
we want to make the reader more sensitive to the difficulty of the problem
via some
motivating examples.

We start with the mono-criterion interval mapping problem of minimizing the
latency. For \FULLHOM and \COMHOM platforms the optimal latency is
achieved by assigning the whole pipeline to the fastest
processor. This is due to the fact that
mapping the whole pipeline onto one single processor minimizes the
communication cost since all communication links have the same
characteristics. Choosing the fastest processor on \COMHOM
platforms ensures the shortest processing time.

\begin{figure}[tbh]
  \centering
  \includegraphics[width=0.38\textwidth]{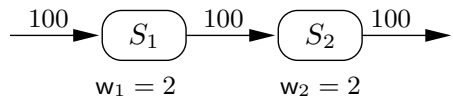}
  \caption{Example optimal with 2 intervals.}
  \label{fig:ex1-pipeline}
\end{figure}

\begin{figure}[tbh]
  \centering
  \includegraphics[width=0.38\textwidth]{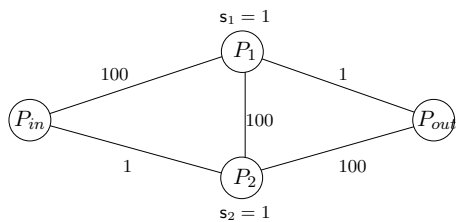}
  \caption{The pipeline has to be split into intervals to achieve an
optimal latency on this platform.}
  \label{fig:L-fullhet}
\end{figure}

However, this line of reasoning does not hold anymore when communications
become heterogeneous. Let us consider for instance the mapping of the
pipeline of Figure~\ref{fig:ex1-pipeline} on the \HETERO platform of
Figure~\ref{fig:L-fullhet}. The pipeline consists of two stages, both
needing the same amount of computation ($\calc = 2$), and the same amount of
communications ($\comm = 100$). In this example, a mapping which minimizes the
latency must map each stage on a different processor, thus splitting
the stages into two intervals.
% illustrates an example for a
%\HETERO platform where the pipeline has to be split into intervals to
%minimize the latency. Let us consider a  %following %homogeneous
%pipeline consisting of 2 stages (see Fig.~\ref{fig:ex1-pipeline}):
%both stages need the same amount of computation of 2 and the
%input size of each stage is the same as its output size, namely 100.
In fact, if we map the whole pipeline on a single processor, we
achieve a latency of $100/100 + (2+2)/1 + 100/1 = 105$,
either if we choose $P_1$ or $P_2$ as target processor.
Splitting the pipeline and hence mapping the first stage on
$P_1$ and the second stage on $P_2$ requires to pay the communication
between $P_1$ and $P_2$ but drastically decreases the latency:
$100/100+2/1+100/100+2/1+100/100=1+2+1+2+1=7$.

Unfortunately these intuitions cannot be generalized when tackling
bi-criteria optimization, where latency should be minimized respecting
a certain failure threshold or the converse.
We will prove in Lemma~\ref{lemma:no_split}
that minimizing the failure probability under a fixed latency
threshold on \FULLHOM and \COMHOM-\FPHOM platforms
still can be done by keeping a single interval. %and hence without
%splitting.

\begin{figure}[tbh]
\centering
\includegraphics[width=0.4\textwidth]{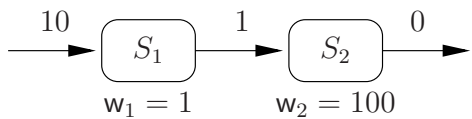}
%\vspace{-0.5cm}
\caption{Example optimal with 2 intervals.}
\label{fig.split}
\end{figure}

However, if we consider \COMHOM-\FPHET, we can find examples in which
this property is not true.
Consider for instance the pipeline of Figure~\ref{fig.split}.
The target platform consists of one processor of speed~$1$ and failure
probability~$0.1$, it is a slow but reliable processor. On the other
hand we have $10$ fast and unreliable processors, of speed~$100$ and
failure probability $0.8$. All communication links have a bandwidth
$\bw=1$. If the latency threshold is fixed to~$22$, the slow processor
cannot be used in the replication scheme. Also, if we use three fast
processors, the latency is $3*10 + 101/100 > 22$. Thus the best
one-interval solution reaches a failure probability of
$(1-(1-0.8^2))=0.64$, which is very high. We can do much better by
using the slow processor on the slow stage, and then replicate ten
times the second stage on the fast processors, achieving a latency of
$10+1/1+10*1+100/100=22$ and a failure probability of
$1-(1-0.1).(1-0.8^{10})<0.2$. Thus the optimal solution does not
consist of a single interval in this case.

%But expanding \COMHOM platforms to \COMHOM-\FPHET, we can find
%counter-examples, where the optimal solution requires the splitting of
%the pipeline into intervals.

%Example Anne----------

\section{Complexity results}
\label{sec:complexity}
In this section, we expose the complexity results for both
mono-criterion and bi-criteria problems. %The polynomial algorithms are
%detailed in Appendix~A, while all optimality proofs are in
%Appendix~B.

\subsection{Mono-criterion problems}

\begin{theorem}
Minimizing the failure probability can be done in polynomial time.
\end{theorem}

\begin{proof}
This can be seen easily from the formula computing the global failure
probability: the minimum is reached by replicating the whole pipeline
as a single interval on all processors. This is true for all platform
types.
\end{proof}

The problem of minimizing the latency is trivially of polynomial time
complexity for \FULLHOM and \COMHOM platforms. However the problem becomes
harder for \HETERO platforms because of the first and last
communications, which should be mapped on fast communicating links to
optimize the latency. Notice that replication can only decrease
latency so we do not consider any replication in this mono-criterion
problem. However, we need to find the best partition of stages into
intervals.

\begin{theorem}\label{th:lat1c}
Minimizing the latency can be done in polynomial time on \COMHOM
platforms.
\end{theorem}

\begin{proof}

The latency is optimized when we suppress all communications. Also,
replication is increasing latency by adding extra communications. On a
\COMHOM platform, the latency is minimized by mapping the whole
pipeline as a single interval on the fastest processor.

\end{proof}

\begin{theorem}\label{th:lat1c:1t1}
Minimizing the latency is NP-hard on \HETERO
platforms for one-to-one mappings.
\end{theorem}

\begin{proof}
The problem clearly belongs to NP. We use a reduction from the
Traveling Salesman Problem (TSP), which is NP-complete~\cite{GareyJohnson}.
Consider an arbitrary instance $\Inst_{1}$ of TSP, i.e., a complete
graph $G=(V,E,c)$, where $c(e)$ is the cost of edge $e$, a source
vertex $s \in V$, a tail vertex $t \in V$, and a bound $K$: is there
an Hamiltonian path in $G$ from $s$ to $t$ whose cost is not greater
than $K$?

We build the following instance $\Inst_{2}$ of the one-to-one latency
minimization problem: we consider an application with $\n=|V|$
identical stages. All application costs are unit costs:
$w_i = \delta_i$ for all $i$. For the platform, in addition to
$P_{\inn}$ and $P_{\out}$
we use $\m = \n =|V|$ identical processors of unit
speed: $s_i = 1$ for all $i$. We simply write $i$ for the processor
$P_i$ that corresponds to vertex $v_i \in V$.

We only play with the link bandwidths: we interconnect $P_{\inn}$ and
$s$, $P_{\out}$ and $t$ with links of bandwidth $1$. We  interconnect
$i$ and $j$ with  a link of bandwidth $\frac{1}{c(e_{i,j})}$.
All the other links are very slow (say their bandwidth is smaller than
$\frac{1}{K+\n+3}$). We ask whether
we can achieve a latency $\latency \leq K'$, where $K' =
K+\n+2$. Clearly, the size of $\Inst_{2}$ is
linear in the size of $\Inst_{1}$.

Because we have as many processors as stages, any solution
to $\Inst_{2}$ will use all processors. We need to map the first stage
on $s$ and the last one on $t$,
otherwise the input/output cost already exceeds $K'$. We spend $2$
time-units for input/output, and $\n$ time-units
for computing (one unit per stage/processor). There remain exactly
$K$ time-units for inter-processor communications,
i.e., for the total cost of the Hamiltonian path that goes from $s$ to
$t$. We cannot use any slow link either.
Hence we have a solution for $\Inst_{2}$ if and only if we have one
for $\Inst_{1}$.
\end{proof}

As far as we know, the complexity is still open for interval mappings,
although we
suspect it might be NP-hard. However, if we relax the interval
constraint, i.e., a set of non-consecutive stages can be assigned to a
same processor, then the problem becomes polynomial. We call such
mappings {\em general mappings}.

\begin{theorem}\label{th:lat1c:gen}
Minimizing the latency is polynomial on \HETERO
platforms for general mappings.
\end{theorem}

\begin{proof}
We consider \HETERO platforms and we want to minimize the
latency.

Let us consider a directed graph with $\n.\m+2$ vertices, and
$(\n-1)\m^2 + 2\m$ edges,
as illustrated in Figure~\ref{fig.lat}.
$V_{i,u}$ corresponds to the mapping of stage~$\S_i$ onto
processor~$P_u$. $V_{0,\inn}$ and $V_{(\n+1),\out}$ represent the
initial and final processors, and data must flow from $V_{0,\inn}$ to
$V_{(\n+1),\out}$.
Edges represent the flow of data from one stage to another, thus
we have $\m^2$ edges for $i=0..\n$, connecting vertex $V_{i,u}$ to
$V_{i+1,v}$ for $u,v=1..\m$ (except for the first and last stages where
there are only $\m$ edges).

\begin{figure}[tbh]
\begin{center}
\includegraphics[width=0.47\textwidth]{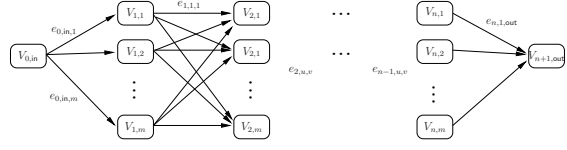}
\end{center}
%\vspace{-0.5cm}
\caption{Minimizing the latency.}
\label{fig.lat}
\end{figure}

Thus, a general mapping can be represented by a path from $V_{0,\inn}$ to
$V_{(\n+1),\out}$: if $V_{i,u}$ is in the path then stage~$\S_i$ is
mapped onto~$P_u$. Notice that a path can create intervals of
non-consecutive stages, thus this mapping is not interval-based.

We assign weights to the edges to ensure that the weight of a path
is the latency of the corresponding mapping. Computation cost of
stage~$\S_i$ on $P_u$ is added on the $\m$ edges exiting $V_{i,u}$,
and thus $e_{i,u,v} = \frac{\calc_i}{\speed_u}$. Communication costs are
added on all edges: %$e(0,\inn,v)= \frac{\com_0}{\bw_{\inn,u}}$ and
$e_{i,u,v}+= \frac{\comm_i}{\bw_{u,v}}$ if $P_u \neq P_v$. Edges
$e_{i,u,u}$ correspond to intra-interval communications, and thus there
is no communication cost to pay.

The mapping which realizes the minimum latency can be obtained by
finding a shortest path in this graph going from $V_{0,\inn}$ to
$V_{(\n+1),\out}$. The graph has polynomial
size and the shortest path can be computed in polynomial
time~\cite{CLRbook}, thus we have the result in polynomial time, which
concludes the proof.% \qedhere

\end{proof}

%% On \COMHOM platforms, we map the whole pipeline as a single interval
%% on the fastest processor. This is not true for \HETERO platforms because
%% of the first and last communications, and the detailed proof for this
%% case is depicted in Appendix~B.

%\section{Minimizing the Failure Probability with Fixed Latency}
%We want to introduce another notation to make the reading of the
%equations in the folloqing section more convenient. We denote $\min
%\speed$ the speed of the slowest processor that is associated to the
%actual interval.

\subsection{Preliminary Lemma for bi-criteria problems}

We start with a preliminary lemma which proves that there is an optimal
solution of both bi-criteria problems consisting of a single interval
for \FULLHOM platforms, and for \COMHOM platforms with identical
failure probabilities.

\begin{lemma}\label{lemma:no_split}
On \FULLHOM and \COMHOM-\FPHOM platforms, there is a mapping of the pipeline
as a single interval which minimizes the failure probability under a
fixed latency threshold, and there is a mapping of the pipeline as a
single interval which  minimizes the latency under a
fixed failure probability threshold.
\end{lemma}

\begin{proof} %{{\bf (Lemma~\ref{lemma:no_split})}}

%As \COMHOM platforms are a generalization of \FULLHOM platforms, we can
%restrict our proof on \COMHOM platforms.

If the stages are split into $p$ intervals, the failure
probability is expressed as
$$1-\prod_{1\leq j \leq p}(1 - \prod_{u\in
\alloc(j)} \fp_u).$$

Let us start with the \FULLHOM case, and with \FPHET for a most
general setting. We can transform the solution into a new one using
a single interval, which improves both latency and failure
probability. Let $k_0$ be the number of times that the first interval
is replicated in the original solution. Then a solution which
replicates the whole interval on the $k_0$ most reliable processors
realizes: (i) a latency which is smaller since we remove the
communications between intervals; (ii) a smaller failure probability
since for the new solution $(1 - \prod_{u\in \alloc(1)} \fp_u)$ is
greater than the same expression in the original solution (the most
reliable processors are used in the new one), and moreover the old
solution even decreases this value by multiplying it by other terms
smaller than~$1$.
Thus the new solution is better for both criteria.

In the case with \COMHOM and \FPHOM, we use a similar reasoning to
transform the solution. We select the interval with the fewest number
of processors, denoted~$k$. In the failure probability expression,
there is a term in $(1-\fp^k)$, and thus the global failure
probability is greater than $1-(1-\fp^k)$ which is obtained by
replicating the whole interval onto $k$ processors.
Since we do not want to increase the latency, we use the fastest $k$
processors, and it is easy to check that this scheme cannot increase
latency ($k \leq k_0$ and the slowest processor is not slower than the
slowest processor of any intervals of the initial solution).
Thus the new solution is better for both criteria, which ends the proof.

%Splitting the pipeline of stages in more than one interval, induces
%the following expression for the failure probability: for $k$
%intervals, we get $$1-\prod_{1\leq j \leq k}(1 - \fp_{P_j}) $$, where
%$\fp_{P_j}$ denotes the failure probability of the processor on wich
%the $j$-th interval is mapped.

We point out that Lemma~\ref{lemma:no_split} cannot be extended to
\COMHOM and \FPHET: instead,
we can build counter examples in which this property is not true, as
illustrated in Section~\ref{sec:examples}.

\end{proof}

\subsection{Bi-criteria problems on \FULLHOM platforms}

For \FULLHOM platforms, we consider that all failure probabilities
are identical, since the platform is made of identical
processors. However, results can easily be extended for
different failure probabilities.
We have seen in Lemma~\ref{lemma:no_split} that the optimal solution
for a bi-criteria mapping on such platforms always consists in mapping
the whole pipeline as a single interval. Otherwise, both latency and
failure probability would be increased.

\begin{theorem}\label{lemma:hom_hom}
On \FULLHOM platforms, the solution to the bi-criteria problem
%an optimal solution minimizing the
%failure probability  under a fixed latency threshold
can be found in
polynomial time using Algorithm~\ref{alg:minFP_hom} or
Algorithm~\ref{alg:minL_hom}.
\end{theorem}

Informally, the algorithms find the maximum number of processors~$k$ that
can be used in the replication set, and the whole interval is mapped
on a set of $k$~identical processors. With different failure
probabilities, the more reliable processors are used.

\begin{algorithm}[bthp]
\label{alg:minFP_hom}
\SetLine
\caption{\FULLHOM platforms: Minimizing $\FP$ for a fixed $\L$}

%procedure {\bf pass1} (node $s\in \NN$)\\
\Begin{

%Algo:
%(if \FPHET, sort processors in non-decreasing order of $\fp_j$) \;
Find $k$ maximum, such that %choose as many processors $k$ to map the whole pipeline onto them such that
$$k\times \frac{\comm_0}{\bw} + \frac{\sum_{1\leq j \leq
n}\calc_j}{\speed} + \frac{\comm_n}{\bw} \leq \L $$
Replicate the whole pipeline as a single interval onto the $k$ (most
reliable) processors\;
}
\end{algorithm}

\begin{algorithm}[bthp]
\label{alg:minL_hom}
\SetLine
\caption{\FULLHOM platforms: Minimizing $\L$ for a fixed $\FP$}

%procedure {\bf pass1} (node $s\in \NN$)\\
\Begin{

%Algo:
%(if \FPHET, sort processors in non-decreasing order of $\fp_j$) \;
Find $k$ minimum, such that %choose as many processors $k$ to map the whole pipeline onto them such that
$$1- (1-\fp^k) \leq \FP$$
Replicate the whole pipeline as a single interval onto the $k$ (most
reliable) processors\;
}
\end{algorithm}

\begin{proof}%{{\bf (Theorem~\ref{lemma:hom_hom})}}

The proof of this theorem is based on Lemma~\ref{lemma:no_split}.
We prove it in the general setting of heterogeneous failure
probabilities. %, thus
% Communications and processors are homogeneous, i.e., $\forall
% i,j,k,l$ $\bw_{\link(i,j) = \bw_{link(k,l)}= \bw}$ and also $\forall
% u,v$ $\speed_u =\speed_v=\speed$. Furthermore the probability to
% crash down is the same for all processors, hence $\fp_u = \fp_v =
% \fp$.
%
An optimal solution can be obtained by mapping the pipeline as a single
interval, thus we need to decide the set of processors~$\alloc$ used for
replication. $|\alloc|$ is the number of processors used.

The first problem can be formally expressed as follows:
%\newpage
\begin{LinearProgram}{\ensuremath 1 - (1- \prod_{u\in \alloc}\fp_{u}) }
%\ensuremath{
 %\EquationsNumbered{ \mark
$$    |\alloc| \frac{\comm_{0}}{\bw} + \frac{\sum_{1\leq i \leq
n}\calc_i}{\speed} + \frac{\comm_n}{\bw} \leq \L
 $$ % }
\end{LinearProgram}

This leads to minimize $\prod_{u\in \alloc}\fp_{u}$,
and the constraint on the latency determines the maximum number $k$ of
processors which can be used:
$$k = \left\lfloor \frac{\bw}{\comm_0} \left(  \L - \frac{\comm_n}{\bw} - \frac{\sum_{1\leq i \leq
n}\calc_i}{\speed} \right) \right\rfloor$$
In order to minimize $\prod_{u\in \alloc}\fp_{u}$, we need to use as
many processors as possible since $\fp_u \leq 1$ for $1\leq u \leq m$.

If one of the most reliable processors is not used, we can exchange it
with a less reliable one, and thus increase the value of the product,
so the formula is minimized when using the $k$ most reliable
processors, which is represented in Algorithm~\ref{alg:minFP_hom}.

% This leads to maximize  $\prod_{1\leq j \leq n}(1- \prod_{1\leq
%  z\leq k_j}\fp_{\alloc(j,z)})$.
%With Lemma~\ref{lemma:no_split}  %As  $\prod_{1\leq j \leq n, n > 1}(1- \prod_{1\leq
%  z\leq k_j}\fp_{\alloc(j,z)}) < 1- \prod_{1\leq z\leq
%  k_j}\fp_{\alloc(1,z)}$
%, we already know that minimize the failure
%  probability, when we map the entire pipeline onto processors without splitting.
%  As the latency is fixed to $\L$, we can choose as many processors $k$
%  as long as $ \sum_{1 \leq j\leq n} k_j\comm_{j-1} + \comm_n \leq \L
%  - \frac{\sum_{1\leq i \leq n}\calc_i}{\speed}$.%%  Finally we choose the $k$
%%   processores with minimal $\fp$, which maximizes  $1- \prod_{1\leq z\leq
%%   k}\fp_{\alloc(1,z)}$.

\medskip

The second problem is expressed below:
\begin{LinearProgram}{\ensuremath   |\alloc| \frac{\comm_{0}}{\bw} + \frac{\sum_{1\leq i \leq
n}\calc_i}{\speed} + \frac{\comm_n}{\bw} }
%\ensuremath{
 %\EquationsNumbered{\mark
$$   1 - (1- \prod_{u\in \alloc}\fp_{u}) \leq \FP
$$  %  }
\end{LinearProgram}

Latency increases when $|\alloc|$ is large, thus we need to find the
smallest number of processors which satisfies constraint~(4). As
before, if one of the most reliable processors is not used, we can
exchange it and improve the reliability without increasing the latency,
which might lead to add fewer processors to the replication set for an
identical reliability. Algorithm~\ref{alg:minL_hom} thus returns the
optimal solution.

\end{proof}

\paragraph{Remark} Both algorithms (\ref{alg:minFP_hom} and
\ref{alg:minL_hom}) are optimal as well in
the case of % if you have a
  heterogeneous failure probabilities. We add the most reliable
processors to the replication scheme (thus increasing latency and
decreasing the failure probability) while $\L$ or $\FP$ are not
reached.
%,It suffices to treat the
%  processors in increasing order of their failure probability. We do
%  not go more into detail, as this platform constellation is not
%  realistic.

%\begin{theorem}\label{lemma:hom_hom_FPfix}
%On \FULLHOM platforms, an optimal solution minimizing the
%latency under a fixed failure probability threshold can be found in
%polynomial time using Algorithm~\ref{alg:minLfixFPfp}.
%\end{theorem}

%In this case, we may be able to reduce the latency by splitting the
%stages in two intervals, at the place where the $\comm_i$ is
%minimum, at a price of a larger failure probability. We prove that we
%can obtain the optimal solution, either by mapping the pipeline as a
%single interval or divided in two intervals.

\subsection{Bi-criteria problems on {\em Com. Homogeneous} platforms}

For \COMHOM platforms, we first consider the simpler case where all
failure probabilities are identical, denoted by \FPHOM.
In this case, the optimal bi-criteria solution still consists of the
mapping of the pipeline as a single interval.

\begin{theorem}\label{lemma:comhom_hom}
On \COMHOM platforms with \FPHOM, the solution to the bi-criteria problem
%an optimal solution minimizing the
%failure probability  under a fixed latency threshold
can be found in
polynomial time using Algorithm~\ref{alg:minFP_comhom_fphom}
or~\ref{alg:minL_comhom_fphom}.
\end{theorem}

Informally, we add the fastest processors to the replication set while
the latency is not exceeded (or until $\FP$ is reached), thus reducing
the failure probability and increasing the latency.
%
%
%\medskip
%

\begin{algorithm}[bthp]
\label{alg:minFP_comhom_fphom}
\SetLine
\caption{\COMHOM platforms - \FPHOM: Minimizing $\FP$ for a fixed $\L$}

%procedure {\bf pass1} (node $s\in \NN$)\\
\Begin{
%Algo:
Order processors in non-increasing order of $\speed_j$\;
Find $k$ maximum, such that
%choose the k fastest processors to map the whole pipeline onto them such that
$$k\times \frac{\comm_0}{\bw} + \frac{\sum_{1\leq j \leq
n} \calc_j}{\speed_k} + \frac{\comm_n}{\bw} \leq \L
$$
Replicate the whole pipeline as a single interval onto the fastest $k$
processors\;
// Note that at any time $\speed_k$ is the speed of \\
// the slowest processor used \\
// in the replication scheme.
 }
 \end{algorithm}

\begin{algorithm}[bthp]
\label{alg:minL_comhom_fphom}
\SetLine
\caption{\COMHOM platforms - \FPHOM: Minimizing $\L$ for a fixed $\FP$}

%procedure {\bf pass1} (node $s\in \NN$)\\
\Begin{
%Algo:
%Order processors in non-decreasing order of $\speed_j$\;
Find $k$ minimum, such that
%choose the k fastest processors to map the whole pipeline onto them such that
$$ 1 - (1-\fp^k) \leq \FP
$$
Replicate the whole pipeline as a single interval onto the fastest $k$
processors\;
 }
 \end{algorithm}

\begin{proof} %{{\bf (Theorem~\ref{lemma:comhom_hom})}}

In this particular setting, Lemma~\ref{lemma:no_split} still applies,
so we restrict to mappings as a single interval, and search for the
optimal set of processors $\alloc$ which should be used.

The first problem is expressed as:

\begin{LinearProgram}{\ensuremath 1 - (1- \fp^{|\alloc|}) }
%\ensuremath{
 %\EquationsNumbered{\mark
   $$ |\alloc| \frac{\comm_{0}}{\bw} + \frac{\sum_{1\leq i \leq
n}\calc_i}{\min_{u\in \alloc} \speed_u} + \frac{\comm_n}{\bw} \leq \L $$
  %}
\end{LinearProgram}

The failure probability is smaller when $|\alloc|$ is large, thus we
need to add as many processors as we can while satisfying the
constraint.  The latency increases when adding more processors, and it
depends of the speed of the slowest processors. Thus, if the
$|\alloc|$ fastest processors are not used, we can exchange a fastest
processor with a used one without increasing
latency. Algorithm~\ref{alg:minFP_comhom_fphom} thus returns an optimal mapping.

The other problem is similar, with the following expression:
\begin{LinearProgram}{\ensuremath  |\alloc| \frac{\comm_{0}}{\bw} + \frac{\sum_{1\leq i \leq
n}\calc_i}{\min_{u\in \alloc} \speed_u} + \frac{\comm_n}{\bw} }
   $$ 1 - (1- \fp^{|\alloc|}) \leq \FP $$
  %}
\end{LinearProgram}

We can thus find the smallest number of processors that should be used
in order to satisfy $\FP$, and then use the fastest processors to
optimize latency, which is done by Algorithm~\ref{alg:minL_comhom_fphom}.

\end{proof}

However, the problem is more complex when we consider different
failure probabilities (\FPHET). It is also more natural
since we have different processors and there is no reason
why they would have the same failure probability. Unfortunately for \FPHET, we can
exhibit for some problem instances an optimal solution in which
the pipeline stages must be divided in several intervals.
The complexity of the problem remains open, but we conjecture it is NP-hard.

%This smart algorithm builds a solution in the form of a single
%interval, but we need to select the processors to use by making a
%tradeoff between their speed and their failure probability. This is
%done by sorting the processors into different classes and selecting
%the most appropriate. However, it is not correct since we can have
%several intervals!
%The reverse problem, i.e., minimizing the latency under a
%fixed failure probability threshold, can be solved: binary search...

\subsection{Bi-criteria problems on \HETERO platforms}

For \HETERO platforms, we restrict to heterogeneous failure
probabilities, which is the most natural case. %While both
%mono-criterion problems have a polynomial complexity,
We prove that the
bi-criteria problems are NP-hard. %% (see Appendix~B).

\begin{theorem}
\label{th:het}
On \HETERO platforms, the bi-criteria (decision problems associated
to the) optimization problems are NP-hard.
\end{theorem}

\begin{proof} %{{\bf (Theorem~\ref{th:het})}}

We consider the following decision problem on \HETERO platforms:
given a failure probability threshold $\FP$ and a latency
threshold $\L$, is there a mapping of failure probability less than
$\FP$ and of latency less than $\L$?
The problem is obviously in NP: given  a
mapping, it is easy to check in polynomial time that it is valid by
computing its failure probability and latency.

To establish the completeness, we use a reduction from
2-PARTITION~\cite{GareyJohnson}.
We consider an instance $\II_1$ of 2-PARTITION:
given $m$ positive integers $a_1, a_2, \ldots, a_m$,
does there exist a subset $I \subset \{1, \ldots, m\}$
such that $\sum_{i \in I} a_i = \sum_{i \notin I} a_i$?
Let $S = \sum_{i=1}^{m} a_i$. %Without loss of
%generality, we can assume
%that all $a_j$ are different and strictly smaller than $S/2$
%(hence $S/a_k > 2$ for all $k$).

We build the following instance $\II_2$ of our problem: the pipeline
is composed of a single stage with $\calc=1$, and the input and output
communication costs  are $\comm_0=\comm_1=1$.
The platform consists in $\m$ processors with
speeds $\speed_j = 1$ and failure probability
$\fp_j = e^{-a_j}$, for $1\leq j \leq m$ (thus $0\leq \fp_j \leq 1$).
Bandwidths are defined as
$\bw_{\inn,j} = 1/a_j$ and $\bw_{j,\out}=1$ for $1\leq j \leq m$.

We ask whether it is possible to
realize a latency of~$S/2 + 2$ and a failure probability of
$e^{-S/2}$. Clearly, the size of $\II_2$ is polynomial
(and even linear) in the size of $\II_1$.
We now show that instance $\II_1$ has a solution if and only if
instance $\II_2$ does.

Suppose first that $\II_1$ has a solution. The solution to $\II_2$
which replicates the stage on the set of processors~$I$ has a
latency of~$S/2+2$, since the first communication requires to sum
$\comm_0/\bw_{\inn,j}$ for all processor~$P_j$ included in the
replication scheme, and then both computation and the final output
require a time~$1$. The failure probability of this solution is
$1- (1-\prod_{j\in I} \fp_j) = e^{-\sum_{j\in I} a_i} = e^{-S/2}$.
Thus we have solved $\II_2$.

On the other hand, if $\II_2$ has a solution, let $I$ be the set of
processors on which the stage is replicated. Because of the latency
constraint,
$$\sum_{j\in I} \frac{1}{\bw_{\inn,j}} + 1 + 1 \leq \frac{S}{2} + 2
.$$
Since $\bw_{\inn,j} = 1/a_j$, this implies that $\sum_{j\in I} a_j
\leq S/2$.
Next we consider the failure probability constraint.
We must have
$$ 1-(1-\prod_{j\in I} \fp_j) \leq e^{-\frac{S}{2}}$$
and thus $e^{-\sum_{j\in I} a_j} \leq e^{-S/2}$,
which forces $\sum_{j\in I} a_j \geq S/2$.
Thus $\sum_{j\in I} a_j = S/2$ and we have a solution to the instance
of 2-PARTITION~$\II_1$, which concludes the proof.

\end{proof}

\section{Related work and conclusion}
\label{sec.conc}
In this paper, we have assessed the complexity of trading between response time and
reliability, which are among the most important criteria for a typical user.
Indeed, in the context of large scale distributed platforms such as clusters or grids, failure
probability becomes a major concern~\cite{FreyFox88,geist02development,duarte06},
and the bi-criteria approach tackled in this paper
enables to provide robust solutions while fulfilling user demands (minimizing latency under
some reliability threshold, or the converse).
We have shown that the more heterogeneity in the target platforms, the more difficult
the problems. In particular, the bi-criteria optimization problem is polynomial for \FULLHOM, NP-hard for \HETERO
and remains an open problem for \COMHOM.

An example of a real world application consisting of a pipeline
workflow can be found in~\cite{rr2008-02}. In this work, we study the
interval mapping of the JPEG encoder pipeline on a cluster of workstations.

Several other bi-criteria optimization problems have been considered
in the literature.
For instance optimizing both latency and throughput is quite natural, as these objectives represent
trade-offs between user expectations and the whole system performance. See~\cite{subhlock-spaa96,c133,c134}
for pipeline graphs and~\cite{Naga07} for general application DAGs.
In the context of embedded systems, energy consumption is another important objective to minimize.
Three-criteria optimization (energy, latency and throughput) is discussed in~\cite{MelhemXu07}.

For large scale distributed platforms such as production grids, throughput is a very important criterion
as it measures the aggregate rate of processing of data, hence the global rate at which execution progresses.
We can envision two types of replication: the first type is to replicate the same computation on different processors,
as in this paper, to increase reliability. The second type is to allocate the processing of different data sets
to different processors (say in a round-robin fashion), in order to increase the throughput. Both replication types can be conducted simultaneously, at the price of more resource consumption.
Our future work will be devoted to the study of the interplay between throughput, latency and reliability,
a very challenging algorithmic problem.

%\clearpage
\bibliographystyle{abbrv}
\bibliography{biblio}

\begin{thebibliography}{10}

\bibitem{abawajy04}
J.~Abawajy.
\newblock Fault-tolerant scheduling policy for grid computing systems.
\newblock In {\em International Parallel and Distributed Processing Symposium
  {IPDPS'2004}}. IEEE Computer Society Press, 2004.

\bibitem{albers01}
S.~Albers and G.~Schmidt.
\newblock Scheduling with unexpected machine breakdowns.
\newblock {\em Discrete Applied Mathematics}, 110(2-3):85--99, 2001.

\bibitem{rr2008-02}
A.~Benoit, H.~Kosch, V.~Rehn-Sonigo, and Y.~Robert.
\newblock {Bi-criteria Pipeline Mappings for Parallel Image Processing}.
\newblock Research Report 2008-02, LIP, ENS Lyon, France, Jan. 2008.
\newblock Available at \url{graal.ens-lyon.fr/~vsonigo/}.

\bibitem{c134}
A.~Benoit, V.~Rehn-Sonigo, and Y.~Robert.
\newblock Multi-criteria scheduling of pipeline workflows.
\newblock In {\em {HeteroPar'2007}: International Conference on Heterogeneous
  Computing, jointly published with {Cluster'2007}}. IEEE Computer Society
  Press, 2007.

\bibitem{c133}
A.~Benoit and Y.~Robert.
\newblock Complexity results for throughput and latency optimization of
  replicated and data-parallel workflows.
\newblock In {\em {HeteroPar'2007}: International Conference on Heterogeneous
  Computing, jointly published with {Cluster'2007}}. IEEE Computer Society
  Press, 2007.

\bibitem{Bhat99efficient}
P.~Bhat, C.~Raghavendra, and V.~Prasanna.
\newblock Efficient collective communication in distributed heterogeneous
  systems.
\newblock In {\em {ICDCS'99} 19th International Conference on Distributed
  Computing Systems}, pages 15--24. IEEE Computer Society Press, 1999.

\bibitem{BhatRagra03}
P.~Bhat, C.~Raghavendra, and V.~Prasanna.
\newblock Efficient collective communication in distributed heterogeneous
  systems.
\newblock {\em Journal of Parallel and Distributed Computing}, 63:251--263,
  2003.

\bibitem{CLRbook}
T.~H. Cormen, C.~E. Leiserson, and R.~L. Rivest.
\newblock {\em Introduction to Algorithms}.
\newblock The MIT Press, 1990.

\bibitem{duarte06}
A.~Duarte, D.~Rexachs, and E.~Luque.
\newblock A distributed scheme for fault-tolerance in large clusters of
  workstations.
\newblock In {\em NIC Series, Vol. 33}, pages 473--480. John von Neumann
  Institute for Computing, Julich, 2006.

\bibitem{FreyFox88}
A.~H. Frey and G.~Fox.
\newblock Problems and approaches for a teraflop processor.
\newblock In {\em Proceedings of the Third Conference on Hypercube Concurrent
  Computers and Applications}, pages 21--25. ACM Press, 1988.

\bibitem{GareyJohnson}
M.~R. Garey and D.~S. Johnson.
\newblock {\em Computers and Intractability, a Guide to the Theory of
  {NP}-Completeness}.
\newblock W.H. Freeman and Company, 1979.

\bibitem{geist02development}
A.~Geist and C.~Engelmann.
\newblock Development of naturally fault tolerant algorithms for computing on
  100,000 processors.
\newblock \url{http://www.csm.ornl.gov/~geist/Lyon2002-geist.pdf}, 2002.

\bibitem{SaifPa2004}
T.~Saif and M.~Parashar.
\newblock Understanding the behavior and performance of non-blocking
  communications in {MPI}.
\newblock In {\em Proceedings of Euro-Par 2004: Parallel Processing}, LNCS
  3149, pages 173--182. Springer, 2004.

\bibitem{ieee-sched}
B.~A. Shirazi, A.~R. Hurson, and K.~M. Kavi.
\newblock {\em Scheduling and load balancing in parallel and distributed
  systems}.
\newblock IEEE Computer Science Press, 1995.

\bibitem{subhlock-ppopp95}
J.~Subhlok and G.~Vondran.
\newblock Optimal mapping of sequences of data parallel tasks.
\newblock In {\em Proc. 5th {ACM} {SIGPLAN} Symposium on Principles and
  Practice of Parallel Programming, PPoPP'95}, pages 134--143. ACM Press, 1995.

\bibitem{subhlock-spaa96}
J.~Subhlok and G.~Vondran.
\newblock Optimal latency-throughput tradeoffs for data parallel pipelines.
\newblock In {\em {ACM} Symposium on Parallel Algorithms and Architectures
  {SPAA'96}}, pages 62--71. ACM Press, 1996.

\bibitem{Tel2000}
G.~Tel.
\newblock {\em Introduction to Distributed Algorithms}.
\newblock Cambridge University Press, 2000.

\bibitem{Naga07}
N.~Vydyanathan, U.~Catalyurek, T.~Kurc, P.~Saddayappan, and J.~Saltz.
\newblock An approach for optimizing latency under throughput constraints for
  application workflows on clusters.
\newblock Research Report OSU-CISRC-1/07-TR03, Ohio State University, Columbus,
  OH, Jan. 2007.
\newblock Available at \url{ftp://ftp.cse.ohio-state.edu/pub/tech-report/2007}.

\bibitem{MelhemXu07}
R.~Xu, R.~Melhem, and D.~Mosse.
\newblock Energy-aware scheduling for streaming applications on chip
  multiprocessors.
\newblock In {\em the 28th IEEE Real-Time System Symposium (RTSS'07), Tucson,
  Arizona}, December 2007.

\end{thebibliography}

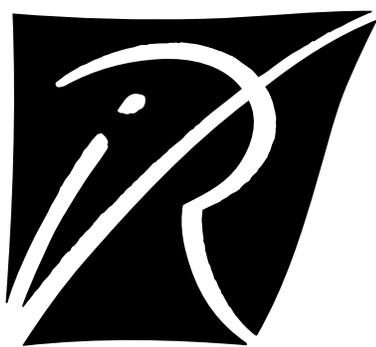
\end{document}